\documentclass[aps,twocolumn, showpacs,preprintnumbers,amsmath,floatfix,longbibliography]{revtex4-1}
\usepackage{graphicx}
\usepackage{dcolumn}
\usepackage{bm}
\usepackage{amssymb}

\begin{document}

\title{Development of magnetic liquid metal suspensions for magnetohydrodynamics}

\author{Florian Carle}
\affiliation{Department of Mechanical Engineering and Materials Science, Yale University, New Haven, CT 06511, USA}
\author{Kunlun Bai}
\affiliation{Department of Mechanical Engineering and Materials Science, Yale University, New Haven, CT 06511, USA}
\author{Joshua Casara}
\affiliation{School of Natural Sciences, University of California, Merced, CA 95343, USA}
\author{Kyle Vanderlick} 
\affiliation{Department of Chemical and Environmental Engineering, Yale University, New Haven, CT 06511, USA}
\author{Eric Brown}
\email{eric.brown@yale.edu}
\affiliation{Department of Mechanical Engineering and Materials Science, Yale University, New Haven, CT 06511, USA}
\affiliation{School of Natural Sciences, University of California, Merced, CA 95343, USA}

\begin{abstract} 
We demonstrate how to suspend various magnetic and non-magnetic particles in liquid metals  and characterize their  properties relevant to magnetohydrodynamics (MHD).   The suspending method uses an acid as a flux  to eliminate oxidation from both metal particles and liquid,  which allows  the particles to be wetted and suspend into the liquid if the particles have higher conductivity than the liquid.  With this process we were able to suspend a wide range of particle materials and sizes from 40 nm to 500 $\mu$m, into three different liquid metal bases,  and volume fractions $\phi$ up to the liquid-solid  transition $\phi_c$.  By controlling the volume fraction of iron particles in liquid eGaIn, we increased the magnetic permeability by a factor of 5.0 and the electrical conductivity by 13\% over that of the pure liquid metal, which gives these materials the potential to exhibit strong MHD effects on the laboratory scale that are usually only observable in the cores of planets and stars. By adding non-magnetic zinc particles, we increased the viscosity by a factor of 160 while keeping the magnetic and electrical properties nearly constant, which would allow independent control of MHD effects from turbulence. We show that the suspensions flow like Newtonian fluids up to the volume fraction  of the liquid-solid transition $\phi_c$.
 
%\noindent \textbf{Keywords}: magnetohydrodynamics, suspensions, magnetic susceptibility, liquid metal, eGaIn, dynamo
\end{abstract}

%47.65.-d  Fluid dynamics: Magnetohydrodynamics and electrohydrodynamics, 
%81.65.Mq Materials science: Oxidation }
% 83.80.Hj rheology: suspensions, 
%81.05.Zx new materials

%47.57.E- fluid dynamics suspensions
%83.85.Cg rheological measurements,
%83.80.Gv magnetorheology
%07.55.Jg magnetometers -susceptibility
%81.16.Be materials science: chemical synthesis methods

\maketitle 

\section{Introduction}

%Definitions , broad problem + broad goal
Magnetohydrodynamic (MHD) phenomena involve the interaction of magnetic fields with fluid flows. For example, many planets and stars exhibit a spontaneous dynamo effect in which a magnetic field is generated by flow of a conducting fluid. This, and other MHD phenomena which depend on fluid flow advecting magnetic fields, occur at large magnetic Reynolds number $Re_m$, which is a characteristic ratio of advection to diffusion of magnetic fields in the flow. $Re_m = \sigma\mu_0(1+\chi)UL$, where \textit{U} and \textit{L} are characteristic velocity and length scales of the system, $\mu_0$ is the permeability of free space, $\chi$ is the magnetic susceptibility of the fluid, and $\sigma$ is its electrical conductivity. A high value of $Re_m$ is easily achieved for the large scale $L$ of planets and stars, but this is much harder to achieve on the smaller scale of a laboratory. Our goal is to develop a material that can easily achieve a large $\chi$ and $Re_m$ on the laboratory scale so that MHD phenomena may be more easily studied experimentally and used for practical applications. 

% motivation
Making MHD phenomena more accessible on the laboratory scale could also lead to development of new devices. For example, the high conductivity and magnetic susceptibility would be desirable in a MHD generator that converts thermal energy to electric energy without moving parts, which can harness the work done by a change in magnetization in an applied magnetic field, and move electric current generated by the dynamo effect through the conductor \cite{popplewell77, rosensweig14}. The ability to rapidly switch on and off the magnetic response via a controlled magnetic field is also highly desired in devices. For example, in magnetorheological dampers, damping increases in response to a magnetic field due to both an increase in effective viscosity from the magnetorheological effect and eddy currents in the conductor \cite{ripper75, carlson98}.

%background -  suitability of liquid metals
  A spontaneous dynamo in an unconstrained flow requires the highest $Re_m (\stackrel{>}{_\sim} 30)$ to achieve among MHD phenomena \cite{monchaux07}, so this $Re_m$ is our ultimate target goal for materials design. To achieve such high $Re_m$, liquid metals and plasmas are traditionally used in experiments because they are the fluids with the highest conductivities $\sigma$. To contain plasmas requires specific magnetic field arrangements and a laminar flow, which are challenges that have not yet been overcome to create a dynamo \cite{khalzov12}. Liquid sodium is the preferred working fluid for dynamo experiments \cite{stieglitz01, gailitis00, monchaux07, zimmerman14} because of its high conductivity ($\sigma = 9.6\cdot10^6$ S/m)\cite{berhanu10}, but can be a challenge to work with because it has a high melting point and explodes on contact with water. Even then, experiments have had to be at least $0.5$ m in size and have taken at least 7 years to create a dynamo \cite{stieglitz01, gailitis00, monchaux07}. Gallium, which has lower conductivity ($\sigma=3.4\cdot10^6$ S/m) than sodium, has been suitable for observing other MHD phenomena such as the magneto-rotational instability which only require $Re_m>1$ \cite{rudiger05}.   A limitation of liquid metals is that at temperatures at which they are liquid, all metals have negligible $\chi \approx 10^{-6}-10^{-5}$, which is the other material parameter that affects $Re_m$.   An increase in $\chi$ by itself is also of interest as forces on the material from magnetic fields are proportional to $\chi$.  We propose to improve on the achievable range of $\chi$ and $Re_m$ by suspending magnetic particles into liquid metals to create a Magnetic Liquid Metal (MLM) with both large susceptibility $\chi$ from the particles and conductivity $\sigma$ from the liquid metals.

% previous approaches
Suspending magnetic particles in liquid metals has been attempted before with limited success. Attempts to suspend pure iron or nickel particles in non-oxidized gallium failed to make suspensions \cite{ito05, cao08, dodbiba11}. On the other hand, nickel \cite{cao08, xiong14} and FeNbVB \cite{dodbiba11} particles coated with silica were suspended in gallium. However, in one of those cases the liquid metal was intentionally oxidized to allow suspending silica-coated particles \cite{xiong14}. In the other cases, we can infer that the liquid metals were oxidized based on the pictures reported which appear dull, rather than shiny like a pure liquid metal \cite{cao08, dodbiba11}. Because oxidized liquid metals have a thin oxidation film on their surface, they have a yield stress like a solid \cite{xu12}, and so do not flow as a Newtonian liquid as would likely be desired for MHD applications. 

Martin \textit{et al.}~\cite{martin00} suspended iron beads of diameter $d = 6.35$ mm into liquid gallium.   They were able to achieve  effective susceptibilities $\chi\approx 3$.  These particles were larger than the capillary length ($\sqrt{\gamma/\rho g} =3.2$ mm for gallium \cite{hardy85}, where $\gamma$ is the surface tension, $\rho$ is the density and $g$ the gravitational constant), which means that they were heavy enough that the stress due to their weight ($\sim \rho g d = 500$ Pa) was larger than the stress from surface tension ($\sim \gamma/d = 100$ Pa). This stress from their weight was also greater than the yield stress $\tau_y$ ($\approx$ 100 Pa) due to the oxide layer \cite{xu12}. Thus, particles could break through the surface to get into the bulk of the liquid metal regardless of surface tension or whether there was an oxidation layer. However, particles this large can break out through the surface just as easily, settle quickly under the effect of gravity, and are not expected to follow fluid flow due to their inertia, properties which may be undesirable in applications. 

Magnetic particles have been successfully suspended into mercury, in which case it was implied that there was no oxidation \cite{linderoth91, popplewell77}. However, mercury is not a desirable liquid due to its toxicity. Furthermore, only very low volume fractions of magnetic particles were suspended in mercury; resulting in up to $\chi = 2\cdot10^{-4}$ for 2\% by weight Fe-Ni-B particles \cite{linderoth91} and $\chi=3\cdot10^{-3}$ for a few percent iron particles coated with tin \cite{popplewell77}, in both cases this was too small to allow a significant increase in $Re_m \propto 1+\chi$.

To address the weaknesses of previous attempts to suspend magnetic particles in liquid metals, we present in this paper a new suspending method that can be generally applied to suspend a variety of different metallic particles into different liquid metals. By removing and preventing oxidation of the metallic particles with a flux (i.e.~an acid), we make the surface wettable by liquid metals which allows them to suspend into the liquid bulk. The flux also prevents and removes any oxide layer from the liquid metal so it behaves mechanically like a simple liquid \cite{xu12}. The suspending method has the advantage that we can suspend both non-Brownian and Brownian particles; the latter is desirable so the particles do not settle or behave inertially. By increasing the volume fraction of magnetic particles suspended up to the liquid-solid transition %$\phi_c$ ($ = 40.5\%$   by volume for iron in eGaIn, for example), 
we are able to tune $\chi$ and thus $Re_m$ over a wide range. We demonstrate this suspending method using liquid gallium and two of its alloys. These liquids have the advantage that they are liquid at or near room temperature, have low vapor pressure (i.e. they are not toxic like mercury \cite{crc14}), and do not explode on contact with water like sodium, making them relatively easy to work with compared to other liquid metals. However, they do have the disadvantage that they corrode some metals, such as aluminum and copper \cite{kamdar1972}. We explain in detail in this article how this suspending process works generally for a variety of materials so it can be easily reproduced and extended in other laboratories.

By adding a third phase of non-magnetic particles into the suspension, we can additionally tune the viscosity $\eta$ of the fluid. This is useful in particular for tuning the Reynolds number $Re = \rho UL/\eta$ (or, equivalently, the magnetic Prandtl number $Pr_m = Re_m/Re$) independently of $Re_m$. Since both $Re$ and $Re_m$ scale with $U$ and $L$, independent control cannot be done over a significant range with liquid metals alone, and as a result both natural and experimental dynamos have always been at high $Re$ and  turbulent \cite{stieglitz01, gailitis00, monchaux07}. Controlling separately the MHD control parameter $Re_m$ and the turbulence control parameter $Re$ will allow investigation of how each separately contributes to MHD phenomena. 

%summary
The remainder of this article is organized as follows. Section \ref{Sec-Material} is a detailed description of how to create MLM, with a process that can apply to a wide range of materials and particle sizes. Section \ref{sec:suspensionseries}  describes the samples characterized in later sections.  Sections \ref{sec:rheology}, \ref{sec:susceptibility}, and \ref{sec:conductivity}, give characterizations of the viscosity, magnetic susceptibility, and conductivity, respectively of suspensions of magnetic and in some cases also non-magnetic particles for a range of volume fractions up to the liquid-solid transition. Section \ref{sec:storage}  describes how well these properties are retained over long periods. Section \ref{Sec-conclusion} is a quantitative discussion of how these material properties can be taken advantage of to reach the extreme parameter ranges required for MHD applications.

\section{Method to produce magnetic liquid metals}
\label{Sec-Material}

%liquid metals
We used several liquid metals as base liquids for the creation of MLM suspensions. The main one used in this study was eGaIn which is an eutectic alloy of gallium and indium \cite{boisbaudran_1885} at the weight ratio 77/23 (melting point $T_{melt}$ = 15$^{\circ}$C, density $\rho$ = 6250 kg/m$^3$, surface tension $\gamma= 624$ mN/m \cite{dickey08}, electrical conductivity $\sigma$ = 3.40$\cdot10^{6}$ S/m \cite{dickey08}). The two metals were mixed at 35$^{\circ}$C and covered with a solution of hydrochloric acid (HCl), which cleans off and prevents oxidation \cite{xu12}.  Despite the fact that indium is a solid at this temperature, the alloying of gallium indium is so favorable that it combines with liquid Gallium into the liquid alloy with stirring by a glass rod.  Pure gallium ($T_{melt}$ = 29.7$^{\circ}$C, $\rho$ = 6095 kg/m$^3$, $\sigma$ = 3.7$\cdot10^{6}$ S/m) and a eutectic alloy of gallium, indium and tin (weight ratio 68.5/21.5/10 respectively, $T_{melt}$ = -19$^{\circ}$C, $\rho$ = 6440 kg/m$^3$, $\sigma$ = 3.46$\cdot10^{6}$ S/m) have also been used as base liquids. The gallium, indium and the eutectic alloy of gallium, indium and tin were purchased from Gallium Source and each have 99.99\% purity.

\begin{figure}
\centering
\includegraphics[width=.475\textwidth]{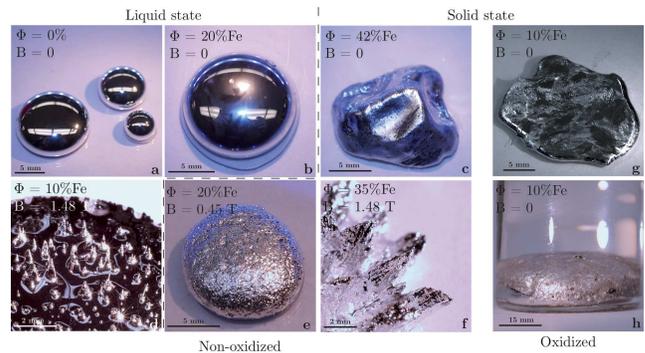}
\caption{Various phenomena observed in a suspension of 29 $\mu$m diameter iron particles in eGaIn, at values of volume fractions $\phi$ and applied constant inhomogeneous magnetic fields with peak magnitude $B$ given in the panels. (a) eGaIn ($\phi=0$). (b) A suspension at $\phi=20\%$, which appeared similar to the pure eGaIn. (c) A yield stress appeared for $\phi$ above the liquid-solid transition $\phi_c=40.5\%$. (d) The Rosensweig instability (e) The magnetorheological effect. (f) The combined effects of panels d and e. The samples in panels a-f were rinsed with HCl to remove oxidation shortly before taking the pictures, and small amounts of remaining HCl are visible in some of the pictures.  The dashed line between panels separates liquid and solid states (i.e. those with zero and non-zero yield stress). Liquid metal states appear shiny because they are conductive and can have smooth surfaces with minimal surface areas due to the surface tension of liquid. Solid states appear rough and  less shiny because the particles poke through the liquid-air interface. They also have a yield stress that prevents them from flowing to obtain a minimal surface area. (g) A sample exposed to air without HCl, which oxidized, causing it to appear dirty and develop a yield stress. (h) A suspension stored for a a few weeks in HCl,  after which samples tended to form a porous solid structure; the liquid state could be recovered by adding more HCl and shaking the sample.}
\label{Fig-Preparation}
\end{figure} 

%hypothesis
When liquid metals are exposed to air, a thin solid oxide layer forms at the surface that appears dirty and dull (Fig.~\ref{Fig-Preparation}g), while the inside remains pure non-oxidized liquid metal protected from air. This oxidized liquid metal wets non-metallic surfaces and does not wet to metals or metal oxides \cite{xu12}.  Indeed, we found that oxidized iron particles which were mixed with eGaIn did not suspend into the bulk -- even  with  stirring vigorous enough for the particles to break the oxide skin on the liquid metal -- and instead stuck to the oxide skin at the surface in equilibrium.  This is true whether or not the eGaIn was recently washed in HCl beforehand so that  the surface appeared shiny and conductive as in Fig.~\ref{Fig-Preparation}a, or if the eGaIn was highly oxidized as in Fig.~\ref{Fig-Preparation}g.  The  inability to wet in equilibrium suggests that the high interfacial tension between liquid metal and oxidized metal particles is a barrier to suspending particles in liquid metals.  For a liquid to wet a solid, a rule of thumb is that the liquid should typically be less polarizable than the solid surface, so that the liquid-solid van der Waals attraction is stronger than the liquid-liquid attraction \cite{fox52, degennes04}. Thus, we propose that using a flux to remove the less-conductive oxide layer ($\sigma =10^3$ S/m) from the iron particles to make the surface more conductive ($\sigma=1.04\cdot10^7$ S/m) and thus more polarizable could enable eGaIn ($\sigma=3.40\cdot10^6$ S/m) to wet the particles (much like soldering) and allow them to suspend. We use an acid solution as the flux, as in a pickling process \cite{hoffman15, hemmelmann13}. Additionally, the acid can remove and prevent oxidation of the liquid metal to prevent a yield stress, allowing the liquid metal to flow like a simple liquid and form into a shape with a minimal surface area \cite{xu12} (Fig.~\ref{Fig-Preparation}a and Movie 1 \cite{supplementary}). 

%procedure
The MLM were made by mixing the liquid metal (typically 2 mL) with varying volume fractions $\phi$ of metallic particles and a HCl solution at room temperature. Enough HCl was used to completely cover the metals so it cleaned off any oxidation and prevented further contact between the metals and air (typically 10 mL HCl for 2 mL of liquid metal). The mixing process is shown in Movie 2 \cite{supplementary} for iron particles ($\rho = 7874$ kg/m$^3$) with mean diameter 29 $\mu$m (the median 90\% range for particle diameters is 18-40 $\mu$m, and there are particles as small as 1 $\mu$m and as large as 219 $\mu$m) suspended in eGaIn at  volume fraction $\phi=10\%$ in a HCl solution with pH $= 0.9$ (the pH meter was calibrated with buffers of pH $=1.68$ and 4). Particles that came into contact with the liquid metal absorbed and suspended into the bulk within a few seconds of contact. Contact could be achieved by initially placing the particles on top of the eGaIn before adding the HCl if the particles were big enough ($\stackrel{>}{_\sim} 1 \mu$m) to settle under gravity. 

%Brownian particles
Using the same process, Brownian particles ($\stackrel{<}{_\sim} 1 \mu$m) instead suspended into the HCl solution (appearing gray).  To suspend Brownian particles, we started with much less HCl, just enough to clean the liquid metal.  Contact between the particles and liquid metal could be achieved by vigorous shaking or stirring of the solution.  For example,  in 30 minutes we were able to suspend up to 4\% iron particles (after which the HCl solution became transparent). We did not attempt further improvements on the process with Brownian particles beyond this proof-of-principle.

\subsection{Understanding the suspending process}
%importance of oxidation
To understand the role of particle oxidation, we can compare the standard experiment described in the previous paragraph with another in which instead initially oxidized iron particles were mixed with initially non-oxidized eGaIn, but before adding HCl.  The eGaIn was rinsed in HCl immediately beforehand to clean off any oxidation, but the HCl was removed before adding the particles.  Without HCl, the particles did not suspend into the bulk of the liquid metal, but instead tended to stick to the surface of the liquid metal.  Once HCl was added  and the oxidation removed \cite{xu12}, the particles absorbed into the bulk of the liquid metal (see Movie 3 \cite{supplementary}). This confirms that the oxidation must be removed from the particles by the flux for suspending to occur, in agreement with the hypothesis that higher conductivity non-oxidized iron surfaces ($\sigma =1.04\cdot10^7$ S/m) are easier to wet and suspend than iron oxide surfaces ($\sigma =10^3$ S/m) \cite{degennes04}.

%importance of wetting
The proposed relation between oxidation  of the liquid metal and suspendability was tested by varying the pH of the HCl solution during the suspending process. For pH $< 0.95\pm0.05$, the surface of the eGaIn appeared shiny, indicating an absence of an oxide layer, and the particles mixed into the bulk of the eGaIn to form a suspension. If instead the HCl solution had pH $> 0.95\pm0.05$, the particles did not suspend into the bulk of the eGaIn even after sitting in contact with eGaIn in the acid bath for 24 hours or stirring the sample for 30 seconds. This critical pH value is consistent with Xu \textit{et al.} \cite{xu12}, who observed that the oxide skin disappeared and wetting of eGaIn to metallic surfaces occurred only for pH $\leq 1 \pm 0.15$. Thus, at higher pH our eGaIn was oxidized, preventing it from wetting the particles. This confirms that oxidation removal from the liquid metal was also required to wet and thus suspend the particles into the liquid.

%\subsection{Chemical reactions}

\begin{figure}
\centering
\includegraphics[width=0.475\textwidth]{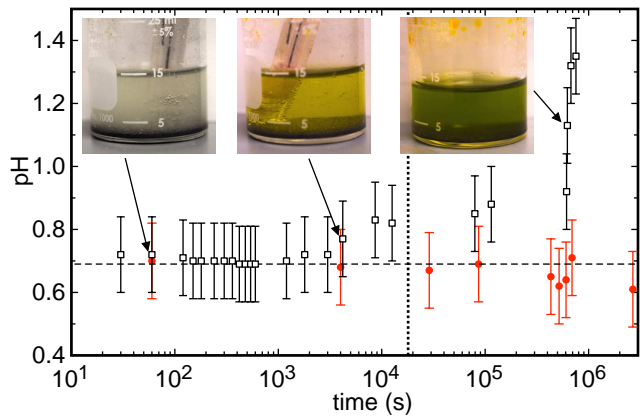}
\caption{pH of samples soaked over long periods of time in HCl. Open squares: iron particles. The pH increased over time, and the solution became green due to the creation of FeCl$_2$ (pictures are  shown at a few times). After $2\cdot10^4$ s of soaking in HCl the particles could no longer be suspended in eGain, indicated by the vertical dotted line.  Solid circles: suspension of $\phi=30\%$ iron particles in eGaIn.  The pH remained constant over 28 days (the dashed line indicates the inital pH), showing that the liquid metal at least partially protected the iron particles from the HCl, allowing particles to be kept in suspension without degradation over long periods of time.}
\label{Fig-pHovertime}
\end{figure}

%oxidation removal
To remove oxidation from the particles in a humid atmosphere, the thermodynamically favored reaction is $Fe_2O_3 + 6 HCl \rightarrow 2 FeCl_3 + 3 H_2O$, where the FeCl$_3$ precipitates into the solution \cite{hoffman15}. Movies 1 and 3 \cite{supplementary} show respectively that enough oxidation removal from the liquid metal and particles happened in matter of seconds to allow the liquid metal to form a minimal surface and wet to non-oxidized metal surfaces \cite{xu12}.

%reaction between iron and HCl
Since the oxidation-removing reaction does not produce the gas bubbles observed in Movie 2 \cite{supplementary}, another reaction must be going on as well. Once the oxidation is removed from the surface of the iron particles exposing the pure iron cores, a reaction that is expected between iron and HCl is 

\begin{equation}
Fe + 2 HCl \rightarrow FeCl_2 + H_2 \ .
\label{Eq-Step2}
\end{equation}

\noindent A flame test confirms that the gas bubbles released are $H_2$ gas. To test the consequence of this reaction, iron particles were placed in an HCl bath (initial pH = 0.69) and the pH was measured over time (Fig.~\ref{Fig-pHovertime}). The error bars of 0.12 on pH represent the standard deviation of repetitions during calibration measurements. As the soaking time increased, the pH increased and the solution turned green, the characteristic color of FeCl$_2$, confirming Eq.~\ref{Eq-Step2}. The bubble production was vigorous over a longer period of time in this experiment than when eGeIn was included in the suspending process, indicating the suspending of iron into eGaIn at least partially protected the iron from the HCl. If the iron particles were soaked in HCl for $\ge 2\cdot10^4$ s before contacting eGaIn, they no longer suspended in eGaIn and the pH  increased slightly. Reducing the bath pH below the critical pH = 0.95 required initially for suspending did not enable these particles to suspend. This suggests that some consequence of the reaction other than the pH change prevented suspending. We hypothesize that the iron particle surfaces were converted to less conductive FeCl$_2$ ($\sigma=10^{-9}$ S/m) from the chemical reaction with HCl (Eq.~\ref{Eq-Step2}), making them less conductive than eGaIn ($\sigma=3.40\cdot 10^{6}$ S/m), thus preventing wetting and suspending. In summary, it appears the bubble producing chemical reaction between iron and HCl did not aid the suspending process, and hindered it if left to go on for several hours. Thus, suspending should occur if the eGaIn, iron particles, and HCl are mixed in any order,  as long as it does not take too long before the  iron particles come in contact with the eGaIn, and the HCl is in contact with the metals for a few seconds to remove any oxidation.

Now that we have a general understanding of the suspending process, we can explain why previous studies failed to suspend uncoated metallic particles into liquid metals \cite{ito05, cao08, dodbiba11}, but could suspend particles with non-metallic coatings into liquid metals \cite{cao08, dodbiba11, xiong14}. The surfaces of the suspensions shown in those articles were not shiny, looking more like Fig.~\ref{Fig-Preparation}g, which is a clear indication of oxidation. The oxide layer prevented the liquid metal from wetting uncoated metal particles, but allowed it to wet nonmetals \cite{xu12}.

\subsection{Robustness of the suspending process} 

\begin{table*}
\caption{The suspending method was demonstrated for a variety of magnetic and non-magnetic metallic particles suspended in eGaIn, listed in the table. The suspensions are ordered by decreasing electrical conductivity $\sigma$, showing that particle materials with higher $\sigma$ than eGaIn ($\sigma = 3.40 \cdot 10^{6}$ S/m, horizontal line), were wetted and suspended, while particle materials with lower $\sigma$ were not wetted or suspended. The particles in bold were those used for quantitative measurements in this paper. References for material source and purity ($\%$): $^a$ U.S. Nano (99.9), $^b$ Alfa Aesar (97.5), $^c$ Chemical Store (99), $^d$ Chemical Store (99.5), $^e$ Chemical Store (97.7), $^f$ Steve Spangler Science (99.5), $^g$ iron was oxidized to obtain an iron oxide surface, $^h$ iron was reacted with HCl to obtain an iron(II) chloride surface.}
\begin{center}
\begin{tabular}{c c c c c c}
 Material & Source & Mean Diameter & Conductivity & Wetted \& \\
 & (purity \%) & ($\mu$m) & (S/m) & Suspended \\
\hline%\vspace{-5pt}\\
\textbf{Zinc} & (see caption) & 0.100$^a$, \textbf{8}$^b$, 500$^c$ & \textbf{1.69$\cdot$10$^7$} & yes \\
Nickel & Chemical Store (99) & 175 & 1.47$\cdot10^7$ & yes \\
\textbf{Iron} & (see caption) & 0.040$^a$, 0.10$^a$, 0.90$^a$, 8$^d$, \textbf{29$^d$}, 45$^e$, 450$^f$ & \textbf{1.04$\cdot$10$^7$} & yes \\
Steel & Chemical Store (99.4) & 175 & 1$\cdot10^7$ & yes \\
\hline
 Stainless Steel & McMaster-Carr &Plate & 1.36$\cdot10^6$ & no \\
Titanium & McMaster-Carr &Plate & 2.40$\cdot10^6$ & no \\
Nichrome & VWR(99) & Wire & 6.67$\cdot10^5$ & no \\
Iron Oxide$^g$ & Chemical Store (99.5) & 29 & $10^{3}$ & no \\
Iron(II) Chloride$^h$ & Chemical Store (99.5) & 29 & $10^{-9}$ & no \\
Soda Lime glass & Mo-Sci Corp. & 125 & $10^{-13}$ & no \\
Polystyrene & Sigma-Aldrich & 200 & $10^{-15}$ & no \\
fused Zirconia Silica & Mo-Sci Corp. & 100 & $10^{-18}$ & no \\
\end{tabular}
\end{center}
\label{Tab:particles}
\end{table*}
%Zinc: 0.100 (0.095-0.0105), 8.0 (6-9), 500 (150-850)
%Iron: 0.040 (0.035-0.045), 0.10 (0.095-0.105), 0.90 (0.75-1.05), 9.0 (5-12), 29 (18-40), 45 (37-53), 450 (300-600)

To demonstrate the robustness of the suspending process for various liquid metal bases, the same 29 $\mu$m iron particles were also successfully suspended using the same process, but with the liquid eGaIn replaced by pure liquid gallium at 50$^{\circ}$C (above its melting point of 29.7$^{\circ}$C) or a eutectic alloy of gallium, indium and tin at room temperature.  Using the same process, we also  successfully suspended particles of various materials and sizes of particles (40 nm to 500 $\mu$m mean diameter) in eGaIn, shown in Table \ref{Tab:particles}. Note that particles with diameter $\stackrel{>}{_\sim} 1$ $\mu$m tend to sediment in suspensions, and will not track fluid flow due to their inertia, features which may not be desirable in many applications. The table shows that particles with a higher conductivity $\sigma$ than the liquid metal base (zinc, nickel, iron, and steel) were wetted and suspended into the bulk, while surfaces with lower $\sigma$  (iron oxide, iron chloride, soda-lime glass, polystyrene, fused zirconia silica) were not wetted or suspended into the bulk. Titanium, nichrome and stainless steel surfaces were found to not be wetted, but we did not obtain particles of those materials to test whether they suspended. These observations confirm the proposal that the particles should be more conductive than the liquid metal to be wetted \cite{degennes04}, and this is both a necessary and sufficient condition for suspending for the materials we tried.

\subsection{Qualitative properties} 

Non-oxidized MLM (Fig.~\ref{Fig-Preparation}b) looked like pure liquid metal (Fig.~\ref{Fig-Preparation}a). It was shiny,  indicating a conductive surface.   The droplet flowed freely if its container was tilted, and the droplet would take a shape with minimal surface area at rest due to the high surface tension of the liquid,  indicating a lack of a yield stress. In contrast, if the liquid metal surface was oxidized, it appeared less shiny and did not form a minimal surface (Fig.~\ref{Fig-Preparation}g).

 When $\phi$ was increased above a critical volume fraction $\phi_c$ corresponding to a liquid-solid transition, i.e.~jamming transition \cite{LN98} ($\phi_c=40.5\%$ for 29 $\mu$m diameter iron particles in eGaIn), the particles became so densely packed that they poked through the liquid air-interface, making the surface appear rough or matte and less shiny, and resulting in a yield stress that allowed the sample to maintain shapes with non-minimal surface areas (Fig.~\ref{Fig-Preparation}c) \cite{brown11}. 
 
 Here we compare the magnetic properties of MLM to other ferrofluids and magnetorheological fluids, in which magnetic particles are usually suspended in oils rather than liquid metals. We exposed suspensions of 29 $\mu$m diameter iron particles in eGaIn to constant inhomogeneous magnetic fields with peak magnitude $B$ at various volume fractions $\phi$. For low $\phi$ and high $B$, MLM behaved as other ferrofluids; the surface deformed into sharp peaks that aligned with the magnetic field lines (Fig.~\ref{Fig-Preparation}d), a consequence of the Rosensweig instability \cite{rosensweig14}. For higher $\phi$ and lower $B$, the MLM exhibited the magnetorheological effect (Fig.~\ref{Fig-Preparation}e): a yield stress due to an induced dipole-dipole interaction between magnetic particles \cite{rabinow48, zukoski93}. For higher volume fraction samples at strong magnetic fields, the ferrofluid and magnetorheological effects could be observed at the same time (Fig.~\ref{Fig-Preparation}f).  The yield stress fluid can still flow to deform under strong enough magnetic forcing, but the shape of the peaks is different than a Newtonian fluid because of the yield stress. The transition from a simple liquid to these magnetic states in response to introducing or removing a magnetic field happens in a fraction of a second and is reversible (Movie 4 \cite{supplementary}). All of these magnetic properties in MLM appear qualitatively similar to other ferrofluids and magnetorheological fluids.

\section{Materials used for characterization}
\label{sec:suspensionseries}

To characterize  the properties of MLM  relevant to MHD, we report in the following sections measurements of  the viscosity $\eta$, magnetic susceptibility $\chi$, and conductivity $\sigma$ of MLM.  As an initial attempt to survey the parameter space of material properties, three series of suspensions with different ratios of magnetic and nonmagnetic particles were measured. In each case we used the same iron particles with mean diameter 29 $\mu$m used primarily for our suspending process. To tune the viscosity $\eta$ independently of the magnetic properties, we additionally suspended zinc particles with mean diameter 8 $\mu$m (nominally 6-9 $\mu$m, $\rho = 7140$ kg/m$^3$). These  particles are highlighted in bold in Table \ref{Tab:particles}. 

The first series, referred to as \textit{$\phi$Fe}, consists of iron particles suspended in eGain at various volume fractions. The two other series have a constant volume fraction of iron particles (10\% and 25\%, respectively) and a variable volume fraction of zinc particles, respectively referred to as \textit{10\%Fe/($\phi$-10\%)Zn} and \textit{25\%Fe/($\phi$-25\%)Zn}. When results are plotted as a function of volume fraction $\phi$ in the following figures, we refer to $\phi$ as the total volume of all solid particle phases divided by the total volume of the sample. For example, for the series \textit{25\%Fe/($\phi$-25\%)Zn}, $\phi=55\%$ means the volume fraction of iron particles is 25\%, and the volume fraction of zinc particles is 30\%.

\section{Rheology}
\label{sec:rheology}

 We performed rheology measurements to determine the conditions under which the suspensions flow like Newtonian liquids, and to obtain the viscosity $\eta$, the main adjustable material parameter in $Re$.

\begin{figure}
\centering
\includegraphics[width=0.45\textwidth]{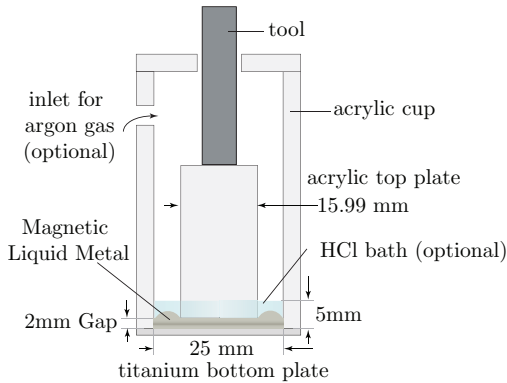}
\caption{Schematic of the modified rheometer set up to measure the viscosity and non-Newtonian properties of liquid metals with a HCl bath or argon atmosphere.}
\label{Fig-Setup}
\end{figure}

Rheology measurements were done using an Anton Paar MCR 302 rheometer in a modified parallel plate geometry \cite{xu12}.  A cross-section of the cylindrical geometry is shown in Fig.~\ref{Fig-Setup}. The suspension was placed in the gap between parallel plates of width $d=2 \pm 0.001$ mm, where the bottom plate is titanium, and the top plate ($D=15.99 \pm 0.02$ mm) is acrylic. Neither surface was corroded or wetted by gallium, eGaIn or the suspensions \cite{xu12}. The suspension extended beyond the outer edges of the top plate (which is unconventional for parallel plate rheology) and was contained in an acrylic cup (diameter $25.10\pm0.05$ mm, height 50 mm) to prevent spillage (a concern for the non-wetting liquids  and large centrifugal forces).  This   containment caused  the suspension to stick  out above the top plate level by about 2 mm, as shown in Fig.~\ref{Fig-Setup}.  The cup also allowed the suspension to be completely covered with a hydrochloric acid (HCl) solution, filled up to 5$\pm$1 mm above the level of the bottom plate to completely cover and prevent oxidation of the liquid metal, which is known to greatly increase the apparent yield stress measured in a rheometer \cite{xu12}. The temperature was maintained at $20.00 \pm 0.01^{\circ}$C by a Peltier element underneath the thermally conductive titanium. 
 
 The torque $T$  on the tool  attached to the top plate was measured while  that tool  was controlled to rotate at constant angular velocity $\omega$.  The mean shear stress was calculated as $\tau = 16T/\pi D^3$ and the mean shear rate at the edge of the plate was calculated as $\dot{\gamma} = D\omega/2d$,  which would correspond to the local field values for a Newtonian fluid in laminar flow in a traditional parallel plate geometry. Since  our measurements do not satisfy these assumptions, these reported $\tau$ and $\dot\gamma$ serve more as  approximate reference scales and do not necessarily correspond to local field values.

 \begin{figure}
\centering
\includegraphics[width=0.475\textwidth]{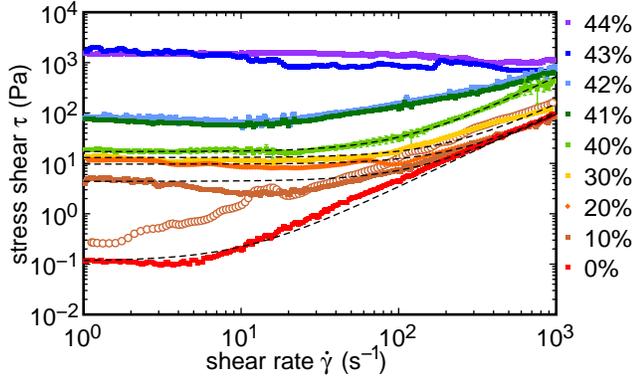}
\caption{Shear stress $\tau$ as a function of shear rate $\dot\gamma$ for suspensions of iron particles in eGaIn (series \textit{$\phi$Fe}). Solid symbols: suspensions in an HCl bath, at various volume fractions $\phi$ shown in the key (upper curves correspond to larger $\phi$). Open symbols: suspension of iron particles in eGaIn at $\phi=10\%$ in an argon atmosphere. Dashed lines: fits of Eq.~\ref{eq-fit} to each $\tau(\dot\gamma)$ curve in a liquid state and an HCl bath. }
\label{Fig-ShearStress}
\end{figure}

The shear stress $\tau$ is shown as a function of shear rate $\dot\gamma$ in Fig.~\ref{Fig-ShearStress} for suspensions of iron particles in eGaIn (series \textit{$\phi$Fe}), for different volume fractions $\phi$. Each curve is the average of three cycles composed of increasing then decreasing logarithmic ramps from $\dot{\gamma}=$ 1 to 1000 s$^{-1}$ with a ramp duration of 60 minutes per decade of $\dot\gamma$, preceded by a pre-shear at $\dot\gamma = 1$ s$^{-1}$ for 60 seconds. The reproducibility of the curves on the repeated cycles confirmed that this pre-shear was enough to eliminate any effects of loading history. The increasing and decreasing ramps did not show any systematic difference (\textit{i.e.}, apparent hysteresis) for liquid states ($\phi<40.5\%$ for series \textit{$\phi$Fe}), indicating that this ramp rate was slow enough to achieve steady state. Each sample exhibited an apparent yield stress, corresponding to the plateau in the limit of small $\dot\gamma$, and a non-linear increase in $\tau(\dot\gamma)$ from which we obtain the viscosity $\eta$ using the methods explained in Sec.~\ref{sec:viscosity}.

\subsection{Yield Stress}

%defining \phi_c
For the series \textit{$\phi$Fe}, samples at $\phi < 40.5\%$ were observed to flow under  the effect of gravity, form into shapes with minimal surfaces, and were shiny (as seen in Fig.~\ref{Fig-Preparation}b), characteristics of a liquid. In contrast, samples at $\phi>40.5\%$ were observed to retain their shapes instead of flow  under gravity (a consequence of their yield stress), and their surface appeared rough or matte due to the particles poking through the liquid-air interface (as seen in Fig.~\ref{Fig-Preparation}c). Some of the samples at $\phi >40.5\%$ fractured underneath the tool during measurements, spilling out of the region beneath the plate and causing a systematic decrease of $\tau_y$  on successive ramps.   Since  these observations are all indicative of a liquid-solid transition, we define it as $\phi_c=40.5\% \pm 0.5\%$ for the series \textit{$\phi$Fe}, where the error represents the difference between $\phi_c$ and the $\phi$ of the nearest measurement point.
 
 \begin{figure}
\centering
\includegraphics[width=0.475\textwidth]{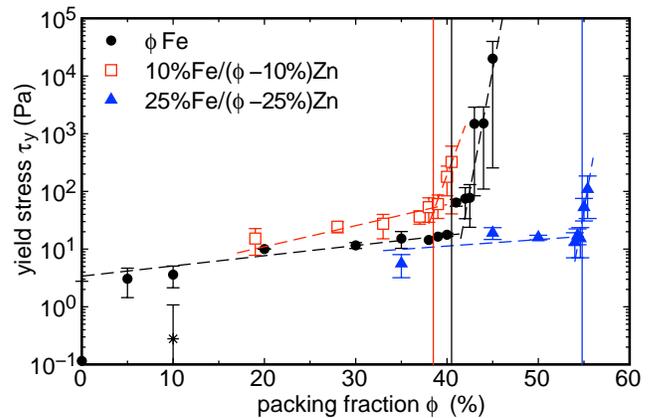}
\caption{Yield stress $\tau_{y}$ of suspensions of various particle mixtures in eGaIn (see key) as a function of volume fraction $\phi$.  Star symbol: suspension  of iron particles in eGaIn at $\phi=10\%$ in an argon atmosphere  instead of an HCl bath. Vertical solid lines:   values of the liquid-solid transition $\phi_c$ for each mixture.  Dashed lines:  separate exponential fits to each of the  liquid and solid regimes for each mixture, which intersect near $\phi_c$.  
}
\label{Fig-YieldStress}
\end{figure}

% yield stress measurement
The yield stress $\tau_y$ was measured as the average stress $\tau$ over the ramps in the range $\dot{\gamma} \le 5$ s$^{-1}$ in Fig.~\ref{Fig-ShearStress}. The yield stress $\tau_y$ is plotted as a function of the volume fraction $\phi$ in Fig.~\ref{Fig-YieldStress}.  The error bars in Fig.~\ref{Fig-YieldStress} were measured as the standard deviation of $\tau$ values used to obtain $\tau_y$.  Figure \ref{Fig-YieldStress} reveals two different scaling regimes of $\tau_y$,  with a sharp increase near $\phi_c$. Fitting the data in each regime by an exponential function with percentage errors as shown in Fig.~\ref{Fig-YieldStress} yields an intersection which is within 1.5\% of $\phi_c$. This  agreement  suggests the sharp increase in yield stress is a result of the liquid-solid transition \cite{brown11}.

%argon atmosphere
The small apparent yield stress $\tau_y$ observed in Fig.~\ref{Fig-ShearStress} for suspensions at $\phi < \phi_c$ is at first glance inconsistent with the conclusion that these states are liquid, since $\tau_y$ should be zero for a liquid. During these experiments, we observed that bubbles appeared at the suspension-acrylic interface, which may be H$_2$ gas as a result of a reaction between iron and HCl (Eq.~\ref{Eq-Step2}). We note that the $\phi=0$ suspension, which has no iron particles in it, had no  resolvable yield stress $\tau_y$ (the measured value is below the rheometer resolution of $6.2\times10^{-2}$ Pa), confirming that the iron particles are needed to produce the apparent yield stress $\tau_y$. If gas bubbles formed in the sample, they could get stuck in between particles or the gap between the plates, which is comparable to the capillary length, which would resist flow with an apparent yield stress $\tau_y$.  Similar trends of increasing  apparent yield stress in $\phi$ up to $\phi_c$ due to increasing volumes of trapped bubbles in suspensions have been observed before \cite{brown11}. To test whether the bubbles were responsible for the apparent yield stress $\tau_y$, we performed an experiment under an argon atmosphere in the flow chamber instead of the HCl bath, which  still prevented oxidation of the liquid metal but did not produce any bubbles.  As experiments under argon atmosphere were more challenging to perform than with the HCl bath, we only measured one sample in an argon atmosphere. The resulting measurement of $\tau_y$ is shown in Fig.~\ref{Fig-ShearStress} for $\phi=10\%$ iron particles.  It is consistent with $\tau_y=0$, and significantly smaller than the value for the same suspension in the HCl bath. This confirms that the apparent yield stress $\tau_y$ is due to bubbles, and not an intrinsic property of the suspension, which behaves like a liquid without yield stress within our resolution for $\phi < \phi_c$.

While trapped bubbles could potentially overwhelm the relatively small hydrodynamic shear stresses in MHD experiments, they would not be expected to become trapped and cause a yield stress in a flow chamber with dimensions large compared to the capillary length.  Figure \ref{Fig-Preparation}b already shows an example of this: the suspension forms a smooth minimal surface  in absence of the confinement of the rheometer plates.  If that  sample had an intrinsic  yield stress of 10 Pa as suggested by Fig.~\ref{Fig-YieldStress}, that yield stress would support rigid protrusions on the surface of the suspension of height $\tau_y/\rho g \sim 0.2$ mm, as in Fig.~\ref{Fig-Preparation}c.

%connection to rest of paper
 Since we desire MLM in a liquid state, measurements of $\eta$, $\chi$ and $\sigma$ in the following sections will be shown for samples only for $\phi< \phi_c$. While the value obtained for $\phi_c$ is lower than expected for non-interacting particles \cite{ohern02}, in practice the liquid-solid transition can appear at much lower $\phi$ if there are interactions between particles \cite{trappe01}. Our value of $\phi_c=40.5\%$ is still far beyond the 2\% by weight achieved by earlier methods of suspending magnetic particles in liquid metals \cite{popplewell77, linderoth91}

\subsection{Viscosity}
\label{sec:viscosity}

\begin{figure}
\centering
\includegraphics[width=0.475\textwidth]{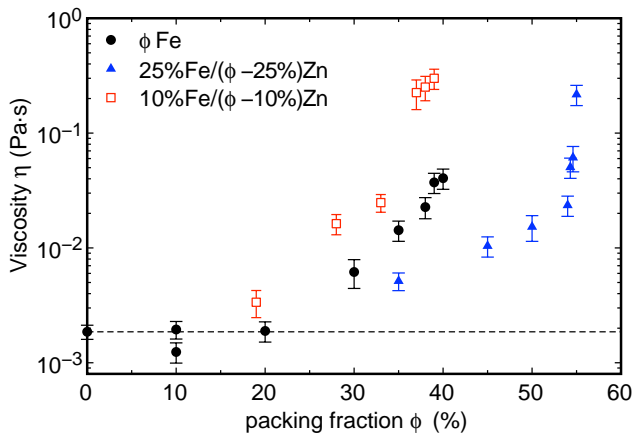}
\caption{Viscosity $\eta$ of suspensions of various mixtures of particles in eGaIn (see key) as a function of volume fraction $\phi$. Horizontal dashed line: $\eta$ for pure eGaIn ($\phi=0$). By tuning the volume fraction of non-magnetic zinc particles at fixed volume fraction of 10\% iron particles (series \textit{10\%Fe/($\phi$-10\%)Zn}), $\eta$ was increased by a factor up to 160 without changing magnetic susceptibiliity $\chi$.}
\label{Fig-Viscosity}
\end{figure}

%obtaining Re_c
Traditionally, the viscosity of suspensions would be obtained from a range where the shear stress $\tau$ is proportional to the shear rate $\dot{\gamma}$  in a  laminar, or low  Reynolds number flow. However, this flow regime where this scaling occurs is not directly accessible for liquid metals in a rheometer measurement due to the low viscosity and the high apparent yield stress $\tau_y$ of the fluids \cite{xu12}. Rather, $\tau(\dot\gamma)$ shown in Fig.~\ref{Fig-ShearStress} increases non-linearly.  We instead use the technique of Xu \textit{et al.} \cite{xu12} to obtain $\eta$ by taking advantage of a hydrodynamic similarity scaling, in which the dimensionless variable $\tau/\eta\dot\gamma$ must be a universal function of Reynolds number Re ($= \rho\dot{\gamma}d^2/\eta$) for different viscosity fluids in the same flow geometry. For the boundary layer-dominated turbulent flow regime, a semi-empirical scaling is known to apply over an intermediate range of Re \cite{xu12}: 
\begin{equation}
\label{eq-fit}
\tau=\tau_{y}+ \left(\frac{\rho\eta\dot{\gamma}^3d^2}{Re_c}\right)^{1/2}
\end{equation}
where $Re_c$ is an unknown critical Reynolds number inherent to the flow geometry. As all the parameters other than $Re_c$ are known for pure eGaIn ($\eta$ = 1.86$\cdot$10$^{-3}$ Pa$\cdot$s \cite{dickey08}), $Re_c$ was obtained by fitting Eq.~\ref{eq-fit} to the averaged ramps shown in Fig.~\ref{Fig-ShearStress} for pure eGaIn ($\phi=0$), with input errors on the stress equal to the 39\% standard deviation of the ramps.  From this we obtain $Re_c = 4.0\pm0.2$ with a reduced chi-squared of  about 1. 

%obtaining eta
Once $Re_c$ was obtained from the calibration with pure eGaIn, then fits of Eq.~\ref{eq-fit} to data  from Fig.~\ref{Fig-ShearStress} with $Re_c$ fixed and $\eta$ as a free parameter were used to obtain the viscosity $\eta$ for suspensions.  As this method relies on an empirical scaling in which we can only confirm the same scaling function holds for a range of Re where we fit for pure eGaIn, we only fit suspension data to the same range of Re ($13.7 < Re < 13700$).
% the highest value for eGaIn (4 for $\dot{\gamma} = 5.9 s^{-1}$ to 680 for $\dot{\gamma} = 1000 s^{-1}$ for $\phi$ = 40\% iron particles)
The corresponding fits of Eq.~\ref{eq-fit} to the averaged ramps shown in Fig.~\ref{Fig-ShearStress} are shown in Fig.~\ref{Fig-ShearStress} for each $\phi<\phi_c$. With an average input error on the stress of 42\% corresponding to the standard deviation of the ramps, the reduced chi-squared ranged from 1-2. The fact that the data for $\phi<\phi_c$ was  fit well by Eq.~\ref{eq-fit} with a reduced chi-squared close to 1 confirms that the macroscopic energy dissipation in the suspensions under shear is consistent with that of a Newtonian fluid for $\phi<\phi_c$.  

The resulting values of $\eta$ are shown in Fig.~\ref{Fig-Viscosity} for different volume fraction $\phi$.  The errors plotted are output errors of the fits of Eq.~\ref{eq-fit}, which had an average of 20\% for the suspensions. The trend of increasing $\eta$ with $\phi$ is typical of suspensions, although the increase does not achieve the divergence at $\phi_c$ of idealized theory \cite{krieger59}, as in practice this divergence tends to be cut off. The viscosity of the suspensions of iron in eGaIn increased up to 22 times higher than eGaIn at $\phi$ = 40\%, the maximum $\phi$ we measured before the liquid-solid transition. 

%fitting a linear function 
Since the viscosities of our suspensions are higher than eGaIn for which this technique was developed \cite{xu12}, we also checked whether we could obtain viscosity values from a linear increase in stress with shear rate as is  traditionally done for a low-Re flow. However, in the relevant range $Re_c < 4$, our measurements are always dominated by the apparent yield stress $\tau_y$ such that $\tau$ is within 8\% of $\tau_y$ even at the highest viscosities reported in Fig.~\ref{Fig-Viscosity},  so there is still no significant range to fit the viscosity to low-Re data \cite{xu12}.

\subsection{Tuning the viscosity with non-magnetic particles}

While the suspensions of iron particles in eGaIn exhibited a significant increase in viscosity $\eta$ on their own, $\eta$ can be tuned independently from the  magnetic susceptibility $\chi$ by adding non-magnetic particles in addition to the magnetic particles already in the MLM suspension. We added zinc particles to fixed volume fractions of iron in each of the two series \textit{10\%Fe/($\phi$-10\%)Zn} and \textit{25\%Fe/($\phi$-25\%)Zn} as defined in Sec.~\ref{sec:suspensionseries}. For each of these series we measured stress $\tau$ vs.~shear rate $\dot\gamma$ as in Fig.~\ref{Fig-ShearStress} (not shown for brevity). The corresponding yield stresses $\tau_y$ are shown in Fig.~\ref{Fig-YieldStress}.  The liquid-solid transition was $\phi_c = 38.5\pm0.5\%$ for series \textit{10\%Fe/($\phi$-10\%)Zn}, and $\phi_c = 54.8\pm0.2\%$ for series \textit{25\%Fe/($\phi$-25\%)Zn},  determined based on observation of the qualitative properties described in Sec.~\ref{sec:viscosity}.  In  each series, exponential functions were fit to $\tau_y(\phi)$  for each of the liquid and solid ranges, which  intersected within 1.5\% of $\phi_c$. We plot $\eta$ in Fig.~\ref{Fig-Viscosity} as a function of volume fraction $\phi$ up to the respective liquid-solid transition $\phi_c$ for each series. In general this indicates a range of parameter space in which $\eta$ can be varied.  Of particular importance is the maximum range we could tune the viscosity $\eta$ by adding zinc particles for a fixed volume fraction of iron particles (so that $\chi$ remains constant). The maximum range we obtained was for the series \textit{10\%Fe/($\phi$-10\%)Zn} where $\eta$ increased up to a factor of 160 before $\phi_c$ was reached.

\subsection{Magnetorheological effect}
\label{sec:MReffect}

\begin{figure}
\centering
\includegraphics[width=0.4\textwidth]{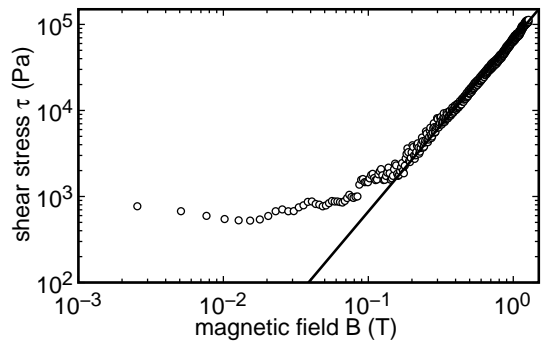}
\caption{Shear stress $\tau$ at fixed shear rate $\dot\gamma$ (approximating a yield stress) as function of a magnetic field $B$, characterizing the magnetorheological (MR) effect for a suspension of $\phi$ = 39\% iron particles in eGaIn (series \textit{$\phi$Fe}).  Solid line: power law fit to the data with exponent 2. Extrapolating the fit to lower $B$ values expected in MHD experiments suggests the yield stress from the MR effect would be negligible compared to hydrodynamic shear stresses. 
}
\label{Fig-MR}
\end{figure}

The magnetorheological (MR) effect results in a yield stress due to an applied magnetic field $B$, as visualized for example in Fig.~\ref{Fig-Preparation}e.
If the MR effect is strong enough, it could in principle produce a yield stress as a result of a self-generated magnetic field from a dynamo. To show an upper bound on the expected MR effect, we used a suspension with $\phi= 39\%$ iron particles in eGaIn (series \textit{$\phi$Fe}), the maximum volume fraction of magnetic particles we could obtain in a liquid state. The suspension was placed in a magnetorheological cell (Anton Paar 1-Tesla Magneto-Rheological Device) with a parallel plate rheometer geometry, and a magnetic field $B$ was applied in the direction of the rheometer tool axis. The sample was sheared at a constant rate $\dot\gamma= 1$ s$^{-1}$  (low enough that the total shear stress was dominated by the yield stress as in Fig.~\ref{Fig-ShearStress}) while the magnetic field $B$ was changed quasistatically in a series of ramps of increasing then decreasing $B$. After the initial ramp, $\tau(B)$ did not exhibit any hysteresis. This approximation of a yield stress is shown as a function of the applied magnetic field $B$ in Fig.~\ref{Fig-MR} for one of the later ramps. 

%resolution limit
For magnetic fields $B<10^{-1}$ T, the shear stress leveled off at a value 2.5  orders of magnitude
%3560\% 
higher than what was measured in the parallel plate setup Fig.~\ref{Fig-Setup}. This can be explained by the fact that when nominally $B=0$ during the ramps, the MR cell still attracted small steel pins, indicating a remnant magnetization of the cell. If we instead demagnetized the cell before loading the suspensions, and measured at $B=0$ before ramping $B$ up, the apparent yield stress was much lower, although still 70\% higher than the value in Fig.~\ref{Fig-ShearStress}. Thus, the plateau value of the apparent yield stress at low $B$ in Fig.~\ref{Fig-MR} is a result of the remnant magnetization of the MR cell, and not an inherent property of the suspension.

%scaling
For magnetic field $B>10^{-1}$ T, the apparent yield stress increased with the applied magnetic field. We fit a power law with exponent 2 -- typical of MR behavior \cite{vicente11} -- to the data for $\tau> 1.46\cdot10^{-3}$ Pa in Fig.~\ref{Fig-MR}. While the MR effect could in principle modify MHD experiments, this effect remained weak at low magnetic field values.  In Sec.~\ref{Sec-conclusion}, we give calculations showing how this would compare to the hydrodynamic shear stress in potential MLM dynamo experiments.

 \section{Magnetic susceptibility}
 \label{sec:susceptibility}

We measured the magnetic susceptibility $\chi$ using a gradiometer \cite{tumanski11}. It consists of two pairs of concentric inductor coils. Suspensions were placed in cylindrical containers of various aspect ratios (defined as length/diameter), and inserted inside one of the inner coils. An alternating current was applied through the outer coils, while an induced voltage was measured in the inner coils. We calculate $\chi$ proportional to the increase in induced voltage over that of an empty coil, corresponding to the effective material response to an applied magnetic field, which is modified from the intrinsic material property by the demagnetization effect \cite{skomski07}. The details of the device and calibrations are presented by Bai \textit{et al.} \cite{bai16}.

\begin{figure}
\centering
\includegraphics[width=0.375\textwidth]{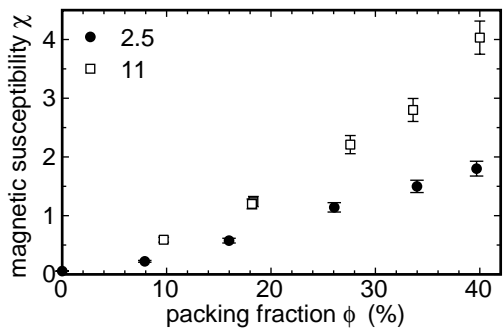} 
\caption{Magnetic susceptibility $\chi$ of suspensions of iron particles in eGaIn (series \textit{$\phi$Fe}) as a function of volume fraction $\phi$, for cylindrical samples with aspect ratios given in the key. By tuning  the volume fraction $\phi$ of magnetic particles, $Re_m\propto 1+\chi$  could increase by a factor up to 5.0 relative to pure eGaIn.}
\label{Fig-Susceptibility}
\end{figure} 

The resulting susceptibility $\chi$ as function of volume fraction $\phi$ is shown in Fig.~\ref{Fig-Susceptibility} for the  series \textit{$\phi$Fe}  at two different sample aspect ratios: 2.5 and 11. The error bars plotted are the quadrature sum of a 7\% systematic uncertainty and a 2.5\% standard deviation of  repetitions \cite{bai16}.  The nearly linear trend of increasing $\chi$ with $\phi$ is expected due to the increase of magnetic material.  However, in detail, $\chi$ is expected to depend in a complicated way on $\phi$, the aspect ratio of the samples, and the aspect ratio of the particles \cite{skomski07}. These dependencies will be reported in detail in \cite{bai16}. As an example to indicate the parameter range achievable, in Fig.~\ref{Fig-Susceptibility}, we obtained up to $\chi=4.0\pm0.3$ for a sample aspect ratio of 11 and $\phi = 40\%$. This results in a potential increase of a factor up to 5.0 in $Re_m \propto 1+\chi$  for MHD experiments over pure eGaIn  at the same conditions, and a much bigger increase in $\chi$ over pure liquid eGaIn ($\chi = 2.19\cdot10^{-6}$) \cite{pashaey73} by a factor up to $1.8\cdot10^6$. 

%zinc independence
One sample of each series \textit{10\%Fe/($\phi$-10\%)Zn} and \textit{25\%Fe/($\phi$-25\%)Zn} was measured and were found to have $\chi$ consistent within the run-to-run variation of the values of series \textit{$\phi$Fe}  with respectively $\phi=10\%$ and $\phi=25\%$ iron.  This  confirms that $\chi$ is  unaffected by the volume fraction of zinc as long  as the volume fraction of iron remains fixed.

\section{Conductivity} 
\label{sec:conductivity}

 We measured the electrical conductivity $\sigma$ of suspensions using the Kelvin sensor method \cite{webster03}.   Suspensions were  first cleaned with HCl, then the HCl  was removed before the suspensions were placed in an acrylic tube with two corrosion-resistant steel electrodes. A different cell was used for each sample series, with radius $r= 3.55 \pm0.01$ mm, and length between the electrodes $L= 45$ with a 2 mm variation between the cells. A direct current $I$ was applied through the length of the sample, while a voltmeter measured the voltage $U$ across the length of the sample with a second pair of electrodes. The conductivity is then given by $\sigma = IL/\pi r^2 U$. We confirmed the measured conductivity was independent of the applied current over the range from 80 to 105 mA with a $\pm10^{-3}$ mA systematic error, which we averaged over to obtain the mean conductivity $\sigma$. The main source of error is the voltage measurement with a systematic error of 1 $\mu$V on typical measurements of 14 to 25 $\mu$V. To test for potential errors from particle sedimentation, the samples were manually shaken at  a frequency of about 2 Hz with an amplitude of 15 cm during the measurements, but by comparison to stationary samples we found the shaking had no significant effect on the measurement.
 
Each cell was first used to measure the conductivity $\sigma$ of pure eGaIn. The values measured were within 1.1\% of the nominal conductivity of eGaIn ($\sigma=3.40\times10^6$ S/m) \cite{dickey08}, which are consistent within the 7\% systematic error on the voltage measurement. This calibration implies the systematic error on following measurements is only 1.1\%.
 
 \begin{figure}
\centering
\includegraphics[width=0.425\textwidth]{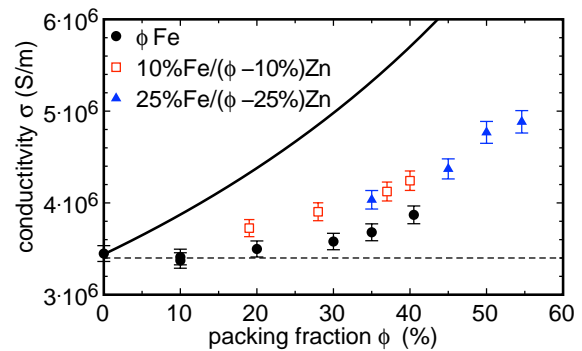}
\caption{Electrical conductivity $\sigma$ as a function of volume fraction $\phi$ for different mixtures of particles in eGaIn indicated in the key. The horizontal dashed line represents the conductivity of the pure liquid metal ($\sigma = 3.40\times10^{6}$ S/m). Solid curve: Meredith \& Tobias model \cite{meredith61} for iron particles in eGaIn (\textit{$\phi$Fe}). The conductivity $\sigma$ only increased over that of pure eGaIn by no more than 42\% for the series \textit{25\%Fe/($\phi$-25\%)Zn}. This means $\chi$ will the main tunable parameter in $Re_m$.} 
\label{Fig-Conductivity}
\end{figure}

% comparison to model
Fig.~\ref{Fig-Conductivity} shows the electrical conductivity $\sigma$ of the three  series  of mixtures as a function of volume fraction $\phi$, up to the liquid-solid transition $\phi_c$ for each series. Each data point on the graph is the average of fifteen  repetitions. The error bars correspond to the sum of the 1.1\% systematic error and a 1.4\% standard deviation of repetitions, which  was the same regardless of whether or not the sample was taken out of the tube and reloaded between measurements. The conductivity $\sigma$ increased by 13\%
%13.8\% , 24.7\%, 41.9\%
over that of eGaIn for the pure iron series \textit{$\phi$Fe}, by 25\% for \textit{10\%Fe/($\phi$-10\%)Zn}, and by 42 \% for \textit{25\%Fe/($\phi$-25\%)Zn}. This confirms that the high  conductivity of the liquid is preserved in the suspensions, and increased in the direction of the higher conductivity particles that were added ($\sigma=1.04\times10^7$ S/m for iron and $\sigma=1.69\times10^7$ S/m for zinc). However, the increase is not as strong as in a simple mixture rule, for comparison.   This increase is also low compared to the model for suspensions by Meredith and Tobias \cite{meredith61}, shown in Fig.~\ref{Fig-Conductivity}. This model  was previously confirmed numerically over the conductivity range of our materials within a scatter of 10\% \cite{bonnecaze91}. Our measured conductivity is systematically lower than the predicted values by up to 35\% in our parameter range, which suggests there might be some opportunity to further improve the conductivity of MLM with different surface treatments.

%  significance for MHD
The measured small increase of the electrical conductivity $\sigma$ of no more than 42\% for series  \textit{25\%Fe/($\phi$-25\%)Zn} compared to the factor of 5.0 increase obtained for $1+\chi$ with increasing volume fraction $\phi$ means the primary way to tune $Re_m$ will be to tune $\chi$ via the volume fraction of magnetic particles.

\section{Retention of properties over time}
\label{sec:storage}

%pH change
To determine how long the samples can retain their properties  under conditions similar  to potential MHD experiments, we stored MLM in HCl and tracked their properties over time.  The pH of a suspension of $\phi=30\%$ iron in eGaIn is shown in Fig.~\ref{Fig-pHovertime}. The pH remained constant within 11\% of the initial pH over 28 days, and the solution remained clear, indicating a lack of FeCl$_2$ formation. This indicates that suspending the iron particles in eGaIn provides some protection against the chemical reaction between iron and HCl (Eq.~\ref{Eq-Step2}). 

%HCl-induced solidification
When MLM were stored in HCl over several weeks without being stirred, the gas pressure in storage containers increased, and the MLM volume expanded with an apparently porous structure shown in Fig.~\ref{Fig-Preparation}h. The MLM also developed a weak yield stress such that the sample would not flow when the container was tilted.  The  tendency for this to occur was stronger with more iron in suspension.  The pressure and porous structure are possibly due to the production of H$_2$ gas from the reaction of Eq.~\ref{Eq-Step2}, while the solidification may be the result of oxidation.  If more HCl was added to lower the pH, and the sample was shaken, the MLM returned to a liquid state  and its original volume.  

For  a suspension of $\phi=30\%$ iron particles in eGaIn that form these solid porous structures after 39 days of storage in HCl, adding HCl  and shaking allowed the suspension to  recover its original viscosity $\eta$, susceptibility $\chi$, and conductivity $\sigma$ within our resolution (20\%, 2.5\%,  and 2.5\%, respectively). After 5 months of storage, the MLM still recovered their original $\eta$ and $\sigma$. However, we observed a 6\% decrease of $\chi_{eff}$ from its original $\chi_{eff}$ (larger than the 2.5\% run-to-run variation).  After 20 months, some samples formed white deposits (possibly due to the formation of FeCl$_2$), and some others rusted. Of those that could be returned to a liquid state, $\chi_{eff}$ was reduced by 80\% compared to its original $\chi_{eff}$. Thus, none of our samples were recoverable after 20 months. 

%removing oxidation
While working with MLM, we frequently ran into situations where the MLM would become oxidized due to exposure to air,  as seen in Fig.~\ref{Fig-Preparation}g, for example when transferring samples between containers. If these oxidized MLM were mixed with HCl, the samples returned to a shiny liquid state, with viscosity $\eta$, susceptibility $\chi$, and conductivity $\sigma$ within resolution of their original values. If the samples were exposed to air for longer periods (up to 1 week), the process of cleaning with HCl took longer, and in some cases we had to change the HCl solution multiple times to return the MLM to a liquid state.

\section{Discussion and conclusions}
\label{Sec-conclusion}
 
In summary, we demonstrate how to make a novel class of materials: magnetic liquid metal (MLM), with both high electrical conductivity $\sigma$ and magnetic susceptibility $\chi$, by suspending magnetic particles in liquid metal. The suspending process is general enough to apply to a wide range of particle materials and diameters from 40 nm to 500 $\mu$m (Table \ref{Tab:particles}), as well as three different liquid metal bases,  and volume fractions $\phi$ up to the liquid-solid  transition (e.g.~$\phi_c = 40.5\%$ for iron particles in eGaIn). The suspending method uses a HCl solution as a flux to eliminate oxidation from both the metal particles and liquid, which allows the liquid metal to wet the particles as long as the particles have higher conductivity $\sigma$ than the liquid. The suspending method  was designed to be easily reproducible in other laboratories as it does not require significant experience in any techniques, and the materials are easily purchased, relatively safe, and can be used at room temperature.

The magnetic susceptibility $\chi$ can be increased by suspending magnetic particles at different volume fractions $\phi$. MHD effects relating to the advection or generation of magnetic fields by fluid flow depend on $Re_m \propto \sigma(1+\chi)$ which was increased by up to a factor of $5.6$ compared to the pure liquid metal;  for series \textit{$\phi$Fe}, $1+\chi$ increased by 5.0 (Fig.~\ref{Fig-Susceptibility}) and $\sigma$ increased by 13\% ($\sigma=3.9\cdot10^6$ S/m) at $\phi$ just below $\phi_c$ (Fig.~\ref{Fig-Conductivity}). This is enough to reach a 2.0 times higher $Re_m$ in eGain-based MLM than liquid sodium ($\sigma = 9.6\cdot10^6$ S/m at a temperature $T=120^{\circ}$C \cite{berhanu10}) at the same values of $U$ and $L$. Furthermore, if these suspending techniques can be applied to liquid sodium, the material properties achieved should allow MHD experiments to reach $Re_m$ about 5 times higher than previous experiments, or alternatively shrinking the volume of the experiment by a factor of $5^3$, which would significantly reduce challenge and cost.  To prevent sodium oxidation without causing an explosion, this would require a non-aqueous flux. 
% particles (or surface coatings) with higher conductivity $\sigma$ than sodium (i.e. copper, aluminum, gold, silver) and -> iron and zinc already meet this criteria
 
 \begin{figure}
\centering
\includegraphics[width=0.475\textwidth]{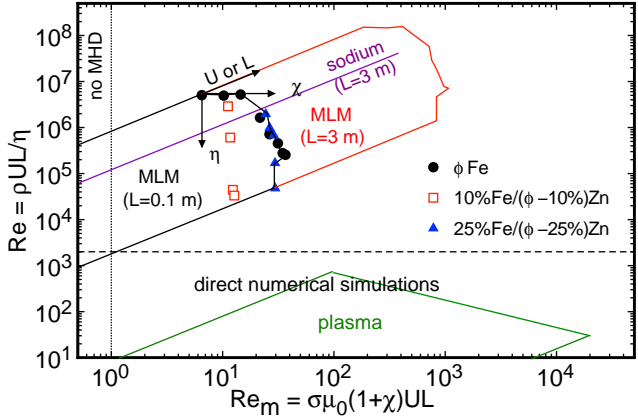}
\caption{(color online) Parameter range accessible for MLM in terms of magnetic Reynolds number \textit{Re$_m$} and Reynolds number \textit{Re}. Symbols: values calculated from the measured values of $\eta$, $\chi$, and $\sigma$ for different $\phi$ from Figs.~\ref{Fig-Viscosity},
\ref{Fig-Susceptibility}, and \ref{Fig-Conductivity} for the sample series listed in the key, and assuming a flow velocity $U = 15$ m/s and $L = 0.1$ m. Region outlined in black: parameter range accessible with the above suspension parameters and controlling flow velocity in the range $U \le 15$ m/s. Region outlined in red: parameter range accessible with the above suspension parameters and controlling $U \le 15$ m/s for $L=3$ m, corresponding to the size of the Maryland sodium facility \cite{zimmerman14}.  Solid line: liquid sodium at 120$^{\circ}$C, $U \le 15$ m/s and $L = 3$ m \cite{zimmerman14}. Region outlined in green: parameter range proposed for plasma  experiments \cite{cooper14}. Direct numerical simulations and plasma experiments are confined to the laminar regime below the dashed line ($Re \stackrel{<}{_\sim} 2000$). For $Re_m \stackrel{<}{_\sim} 1$ (dotted line), no significant MHD effects are expected. The arrows represent the direction that can be moved in the parameter space by tuning the corresponding labeled variable. MLM  can be tuned to access both dimensions of the parameter space, spanning both turbulent and laminar regimes,  and reaching a higher $Re_m$ than sodium  for the same dimensions.}
\label{Fig-Ratio}
\end{figure}

The viscosity $\eta$ can be increased independently of the magnetic susceptibility $\chi$ at a fixed volume fraction of magnetic particles by additionally suspending varying volume fractions of non-magnetic particles. By approaching the liquid-solid transition $\phi_c$, we were able to increase $\eta$ by factor of 160 for series \textit{10\%Fe/($\phi$-10\%)Zn} (Fig.~\ref{Fig-Viscosity}). 

The parameter space in terms of $Re_m$ and $Re$ that could be achieved based on our measured range of $\chi$, $\sigma$, and $\eta$, and calculated $\rho$ based on the densities and volume fractions of materials is shown in Fig.~\ref{Fig-Ratio}. A modest experiment scale of $L=0.1$ m and $U \le 15$ m/s is large enough to reach $Re_m=37$, above the threshold for a spontaneous dynamo ($Re_m=30$)\cite{monchaux07}. Fixed material properties (\textit{i.e.}, $\chi$, $\sigma$, $\eta$, and $\rho$) only allow movement along a diagonal line of slope 1 in the parameter space by varying $U$ or $L$, as is the case for liquid sodium  at a temperature of 120 $^{\circ}$C shown in Fig.~\ref{Fig-Ratio}.   On the other hand, the ability to independently control the material properties by suspending different particles allows access to both dimensions of the parameter space. The material properties of the fluid in MHD are typically characterized by the magnetic Prandtl number ($Pr_m = Re_m/Re$) which can only be varied slightly  for pure liquids by changing temperature. With MLM, we achieved a range from $Pr_m= 1.6\cdot 10^{-6}$ for pure eGaIn, up to $Pr_m=6.2\cdot10^{-4}$ (a factor of 440) just below the liquid-solid transition for series \textit{25\%Fe/($\phi$-25\%)Zn}.

% spanning laminar to turbulent regime
The existing approaches to MHD cannot span the laminar-to-turbulent transition; liquid metal experiments are always turbulent, and direct numerical simulations and plasma experiments are confined to the laminar regime due to computational cost and flow instabilities, respectively.  For example, the parameter range of proposed helium plasma experiments  with temperature $T=1.0\cdot10^4 - 5.6\cdot10^5$$^{\circ}$C, $U\le 10$ km/s and $L=3$ m \cite{cooper14} are shown in Fig.~\ref{Fig-Ratio}.
 Figure \ref{Fig-Ratio} suggests that MLM could span could span from the turbulent to the laminar regimes of $Re$ as a result of the ability to increase the viscosity $\eta$.    For example, $Re \approx 1600$ could be reached for the highest $Pr_m$ we found (corresponding to $\phi$=55\% for the series \textit{25\%Fe/($\phi$-25\%)Zn}) while still achieving $Re_m \ge 1$ (the minimum threshold for most MHD phenomena), assuming a fixed $\chi=2.2$ , $\rho=6923$ kg/m$^3$, $\sigma=4.9\cdot10^6$ S/m and $\eta = 0.2$ Pa$\cdot$s. Thus, an experiment carefully designed to take advantage of this small available laminar parameter range in Fig.~\ref{Fig-Ratio} would allow the first approach to MHD that could span between the turbulent and laminar regimes and connect to the parameter regime of  direct numerical simulations.

%MR effect
A self generated magnetic field in a dynamo could in principle generate a MR effect in MLM. If we extrapolate the fit of the data in Fig.~\ref{Fig-MR} down to a magnetic field of $B=4\cdot10^{-3}$ T (the value measured in the Cadarache dynamo \cite{monchaux07}), the yield stress induced by the magnetic field would be $\tau_y\approx  0.7$ Pa,  negligible compared to the hydrodynamic shear stress given by the latter term of Eq.~\ref{eq-fit} $\tau = (\eta^2 Re^{3/2})/(\rho L^2 Re_c^{1/2}) \approx 22$ Pa for the lowest stress experiments we propose, i.e.~laminar MHD  at $Re\approx1600$, where $\eta = 0.2$ Pa$\cdot$s, $\rho$ = 6923 kg/m$^3$ and L = 0.1 m. Therefore, if the magnetic field is self-generated as in a Cadarache-type dynamo and not applied by an external source, it is expected that the samples will remain in a liquid state during the experiments without a significant MR effect.

%lorentz force
The other aspect of MHD is the Lorentz force on fluid flow due to magnetic fields, which also increases with $\chi$. One source is a dynamo effect which creates an induced magnetic field which applies a Lorentz force on the induced current. While this feedback mechanism exists in astrophysical MHD, no laboratory experiment has reached this regime yet. This regime could be achieved if the ratio of Lorentz stress to inertial stress $N^*=2\sigma(1+\chi)^2H^2 L/f\rho U$  becomes comparable to one.  We fix the parameters values $L=0.1$ m, $f\approx 0.02$ as a typical value for a friction factor \cite{schlichting03}, and an induced magnetic field of $H=3 \cdot10^{3}$ A/m (the value obtained in the Cadarache dynamo at the threshold $Re_m = 30$ \cite{monchaux07}). Since $N^*$ decreases with $U$, and $Re_m$ increases with $U$, we can adjust $U$ to find the value of $N^*$ at which the minimum $Re_m=30$  for a dynamo can still be obtained for different materials.   We obtain $N^*=0.07$ for liquid sodium ($\sigma=9.6\cdot10^6$ S/m, $\rho=927$ kg/m$^3$), $N^*=0.2$ for our suspensions of $\phi=40\%$ iron in eGaIn  (series \textit{$\phi$Fe}, with $\chi=4.0$, $\sigma=3.9\cdot10^6$ S/m, $\rho=6900$ kg/m$^3$), and $N^*=2$ for a hypothetical suspension of liquid sodium with $\phi=40\%$ iron particles (assuming $\chi=4$, $\sigma=9.6\cdot10^6$ S/m, $\rho = 3700$ kg/m$^3$). This suggests that MLM with eGaIn could have Lorentz forces with a noticeable effect on the flow, while Lorentz forces could potentially be dominant for MLM using sodium.
 
The tunable range of  viscosity $\eta$, conductivity $\sigma$, and relative permeability $1+\chi$ for the materials reported are expected to be typical of MLM,  but with some opportunity for improvement. The independently tunable range of $\eta$ is expected to be insensitive to material, rather it depends on how far $\phi$ is from the liquid-solid transition $\phi_c$ \cite{krieger59, brown11}. There is potential that $\eta$ could be increased beyond the factor of 160 we obtained; for example, a 3.5 decade increase of $\eta$ with $\phi$ has been achieved with non-MLM suspensions while remaining Newtonian \cite{cheng02}. Since our measured $\sigma$ is systematically lower than predicted values by up to 35\% in our parameter range \cite{meredith61}, there might be some opportunity to further improve the conductivity with different surface treatments.  The increase in $\chi$ with magnetic particles was also predicted to be insensitive to the material as long as it is ferromagnetic \cite{skomski07}, so we expect increases in $Re_m$ of order 1-10 for nearly spherical particles. It also remains to be seen if $\chi$ will retain the higher values  achieved with large sample aspect ratios (Fig.~\ref{Fig-Susceptibility}) or attain the even larger values predicted for larger particle and sample aspect ratios \cite{skomski07} in a turbulent flow where the magnetic field is not uniform or aligned with the sample or particles. If so, extreme sample aspect ratios could be taken advantage of in specific experiments to achieve much higher $\chi$.

\begin{acknowledgments}
The authors acknowledge Marcus Bell and Christopher Bush for their help in exploring suspending procedures.  We acknowledge the financial support from NSF Grant No.~CBET-1255541 and AFOSR grant FA 9550-14-1-0337.
\end{acknowledgments}

\end{document}